\documentclass[a4paper,9pt, twocolumn]{article}
\usepackage{graphicx}
\usepackage[a4paper, total={17.8cm, 25cm}]{geometry}
\usepackage{authblk}
\usepackage[backend=biber, giveninits=true, dateabbrev=false, urldate=long, style=numeric, maxbibnames=99, isbn=false, doi=false, sorting=none]{biblatex}
\bibliography{bibliography}
\usepackage[font=small]{caption}

\title{Underestimation of systolic pressure in cuff-based blood pressure measurement}
\author[1,a]{Kate Bassil}
\author[1,b]{Anurag Agarwal}
\affil[1]{Department of Engineering, University of Cambridge, Trumpington Street, Cambridge, CB2 1PZ, UK}
\affil[a]{ K.B. designed and performed the research, analysed data, and wrote the manuscript}
\affil[b]{A.A. supervised the research, contributed to study design, data analysis, and manuscript writing}
\date{}

\begin{document}

\twocolumn[
  \begin{@twocolumnfalse}
  \maketitle
  \centering
  \vspace{-0.8cm}
  Manuscript compiled \today\\
Correspondence to kjb83@cantab.ac.uk
    \begin{abstract}
      High blood pressure (hypertension) is the number one risk factor for premature death. Hypertension is asymptomatic, so blood pressure must be regularly monitored to diagnose it. In auscultatory blood pressure measurement, a patient’s systolic (maximum) and diastolic (minimum) blood pressure are inferred from the pressure in an inflatable cuff wrapped around the arm. This technique is the gold standard against which all other non-invasive devices are validated. However, auscultatory measurements systematically underestimate systolic blood pressure and overestimate diastolic blood pressure. Overestimation is attributed to the increased cuff pressure needed to occlude the artery because of the surrounding tissue and arterial stiffness. In contrast, the cause of systolic underestimation, which leads to potentially a third of systolic hypertension cases being missed, has remained unclear. When the cuff is inflated beyond the systolic blood pressure, the blood flow to the vessels downstream of the cuff is cut off. The pressure in these downstream vessels drops to a low plateau. We have developed a novel experimental rig that shows that the low downstream pressure is the key cause of the underestimation of systolic blood pressure. The lower the downstream pressure, the greater the underestimation. Our results yield a simple physical model for the underestimation of systolic pressure in our rig and in the human body. Understanding the physics behind the underestimation of systolic blood pressure paves the way for developing strategies to mitigate this error.\\
      \vspace{0.2cm}

\noindent \textbf{Significance Statement:} The cuff-based auscultatory method is the gold standard for blood pressure measurement, but it remains inaccurate. While the overestimation of diastolic (minimum) blood pressure is well understood, the consistent underestimation of systolic (maximum) blood pressure remains unexplained.  Previous experimental setups have neither observed nor investigated this underestimation. With a novel experimental model, we provide an explanation for the underestimation of systolic blood pressure and a better understanding of the physics behind auscultatory measurement. Our findings could lead to improved calibration methods, enhancing the accuracy of blood pressure measurements.\\
\vspace{0.5cm}
    
    \end{abstract}
    
  \end{@twocolumnfalse}
]

\section*{Introduction}

The measurement we refer to as ‘blood pressure’ is not a constant value but oscillates between a maximum (systolic) and minimum (diastolic) value with each contraction and relaxation of the heart (a cardiac cycle). Systolic and diastolic blood pressure are critical measures of human health. Hypertension, the condition of persistently raised blood pressure, is the number one risk factor for premature death worldwide \cite{natureleadingcause,worldhearfed}. It is also typically asymptomatic, so it is unlikely to be apparent until the resulting health problems become significant. Early identification of an asymptomatic condition requires regular and accurate testing.

For regular testing across a large portion of the population, low-cost, widely accessible and non-invasive testing methods are essential. Direct blood pressure measurement with an arterial line can be accurate but is expensive and invasive \cite{hignett2006radial}. Indirect measurement techniques are therefore used for routine blood pressure testing. The gold standard amongst non-invasive blood pressure measurement techniques, against which other devices are validated, is cuff-based auscultatory measurement. For example, new non-invasive measurement devices should give mean results within $5$ mmHg of the auscultatory readings for both systolic and diastolic pressures \cite{stergiou2018universal}. Because of these validation requirements, improvements in auscultatory measurement will translate to greater accuracy in other measurement techniques, such as the widely used automatic blood pressure monitors. 

\begin{figure*}[t]
\centering
\includegraphics[width=17.8cm]{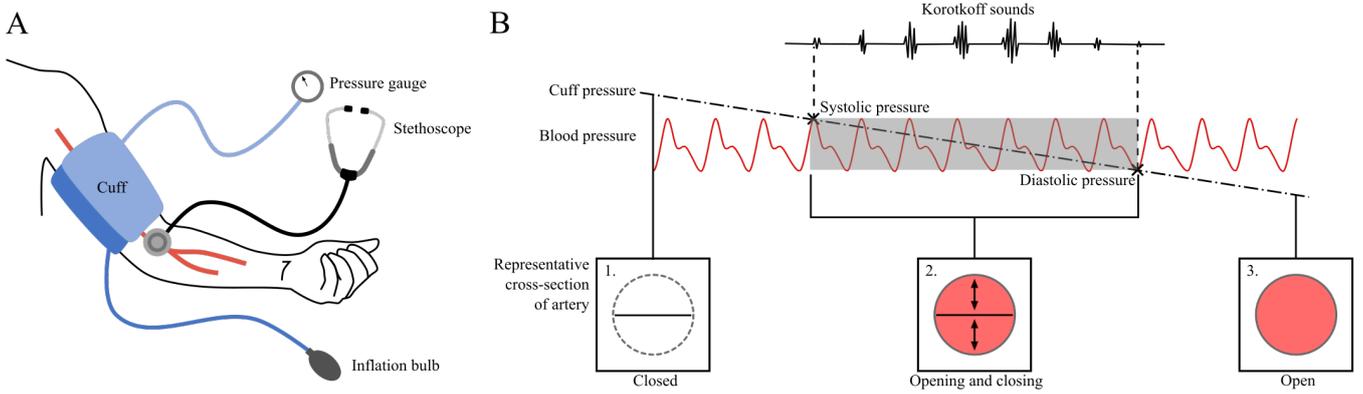}
\caption{A) Setup for auscultatory blood pressure measurement. B) Representative artery behaviour during auscultatory measurement.}\label{fig:auscmethod}
\end{figure*}

\subsection*{The Auscultatory Method}
In the auscultatory method, an inflatable cuff is placed around the upper arm, as shown in Fig.\ \ref{fig:auscmethod}A. The cuff is inflated to a pressure above the systolic pressure. The pressure in the cuff is then gradually released at $2$--$3$ mmHg/s \cite{ogedegbe_2010}. At a critical cuff pressure, periodic tapping sounds known as Korotkoff sounds start, and can be heard through a stethoscope. This critical pressure is recorded as the systolic blood pressure (SBP). As the pressure is reduced below the SBP, Korotkoff sounds are continuously heard until a second critical pressure is reached, when these sounds stop. This cuff pressure is recorded as the diastolic blood pressure (DBP).

Fig.\ \ref{fig:auscmethod}B illustrates the relationship between the cuff pressure, blood pressure, artery cross-section and Korotkoff sounds. With the cuff pressure initially raised above the systolic pressure, the artery is occluded \cite{tavel1969korotkoff} and remains fully closed throughout the cardiac cycle, as shown in cross-section 1 of Fig.\ \ref{fig:auscmethod}B. The cuff pressure is then gradually decreased and drops below the SBP. In the range SBP $>$ cuff pressure $>$ DBP, shaded in grey in Fig.\ \ref{fig:auscmethod}B, the pressure inside the artery is higher than the applied cuff pressure during a portion of each cardiac cycle. For this period of the cycle, the artery can open, and blood can flow through it. For the remainder of the cycle, when the blood pressure drops back below the cuff pressure, the artery is closed. Initially, when the cuff pressure is only slightly below the SBP, the artery opens only for a brief portion of each cardiac cycle. As the cuff pressure decreases further, the artery remains open for a progressively larger portion of each cycle. When the cuff pressure falls below the DBP, the external pressure no longer surpasses the internal pressure at any point in the cycle, allowing the artery to remain fully open, as shown in cross-section 3 of Fig.\ \ref{fig:auscmethod}B. The opening and closing of the artery in the range SBP $>$ cuff pressure $>$ DBP produces Korotkoff sounds \cite{Korotkoff, drzewiecki1989krotkoff, Bloodpressurebook, baranger2023fundamental}. Outside of this range, the artery is either fully closed or fully open, and there is silence \cite{Bloodpressurebook}.

\subsection*{Inaccuracy of cuff-based measurement}

The description above is an idealised version of auscultatory measurement. If the artery closed as soon as the cuff pressure exceeded the internal pressure, as described, then the measured cuff pressure would be a perfect proxy for the arterial blood pressure.

However, in practice, the auscultatory method is inaccurate, systematically underestimating systolic pressure and overestimating diastolic pressure. Meta-analysis of 74 studies by Picone et al. \cite{picone} shows that cuff-based blood pressure measurements underestimate SBP by an average of $5.7$ mmHg and overestimate DBP values by an average of $5.5$ mmHg. The diagnostic impact of these errors is substantial: Turner et al. \cite{turner_2004} found that systematically overestimating DBP by $5$ mmHg increases the number of patients with a measured DBP over $90$ mmHg, the threshold for hypertension, by $132$\%. Equivalent underestimation of SBP results in $30$\% of patients with systolic hypertension (SBP $>140$ mmHg) being missed.

The physical factors contributing to overestimation are well established. Firstly, the cuff pressure is not applied directly to the artery, but to the surface of the arm. The arm tissue reduces the pressure transmitted to the artery, so the external load experienced by the artery is lower than the cuff pressure \cite{Deng_Liang, lan_2011, hargens_1987}. Secondly, additional pressure is required to overcome the buckling stiffness of the artery wall. Therefore, the artery will only close once the external pressure exceeds the internal pressure by a certain threshold, known as the buckling pressure \cite{fung_2013}. These factors cause the measured cuff pressure to overestimate the blood pressure \cite{Deng_Liang, Ma_stiffness}.

Given these existing explanations for inaccuracy, it might be expected that the auscultatory method would overestimate both SBP and DBP. The surrounding tissue and artery stiffness are predicted to increase the cuff pressure required to keep the artery closed and so to cause the artery to reopen early during cuff deflation, while cuff pressure still exceeds SBP. However, intra-arterial pressure measurements taken by Celler et al. \cite{celler_2021} during cuff deflation show that, rather than the artery opening early, the reopening of the artery is substantially delayed. In these measurements, the brachial artery remains closed down to cuff pressures as much as 24 mmHg below the systolic pressure \cite{celler_2021}. There is currently no physical explanation for this delayed opening and the underestimation of SBP it produces. This work aims to identify the cause of underestimation of SBP in auscultatory measurement, addressing a critical gap in our understanding of cuff-based blood pressure measurement.

\begin{figure}[tbhp]
\centering
\includegraphics[width=8.7cm]{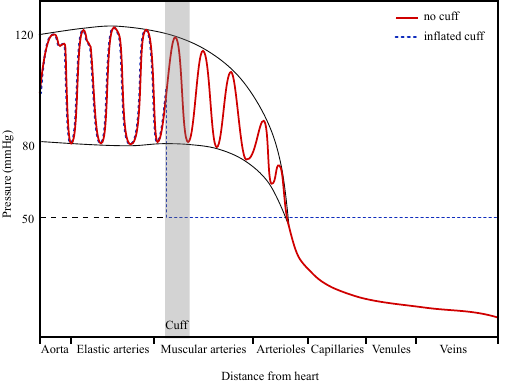}
\caption{Variation of pressure throughout the blood vessels. The red trace shows pressure variation when the vessels are undisturbed, with no cuff applied (adapted from \cite{BPvariation}). The black envelope around the trace shows the SBP and DBP. The blue (dotted) trace shows the blood pressure variation when the brachial artery is closed by an inflatable cuff around the upper arm. This plot shows the steady state condition once the pressure has equalised in the downstream vessels (on initial closure, blood will flow from the higher pressure arteries to the lower pressure veins until the pressures level out \cite{nitzan2005effects}).}
\label{fig:pressurevariationBODY}
\end{figure}

\subsection*{Underestimation of systolic pressure}
To explain underestimation in auscultatory measurement, we propose a new model that takes a more global view of the circulatory network, accounting for the variation of blood pressure along the length of the artery. A schematic of the pressure variation throughout the circulatory system is shown in Fig.\ \ref{fig:pressurevariationBODY}. The variation in blood pressure when the arm is undisturbed, before the blood pressure cuff is applied, is shown in red. The pressure decreases with distance from the heart \cite{BPvariation}, continuing to drop in the smaller arteries, capillaries and veins, where the pressure approaches $0$ mmHg.

When an inflated cuff occludes the artery, the blood flow to the vessels distal to the cuff (downstream of the point of closure) is cut off. In the absence of flow, the blood in the downstream network of vessels becomes static and the pressure in the vessels reaches an equilibrium. This means that the arterial pressure directly distal to the cuff drops towards the pressure of the rest of the downstream vessels, eventually plateauing at a low, constant pressure usually between $30$--$70$ mmHg \cite{tavel1969korotkoff}.  With the downstream vessels separated from the heart, there are no pressure oscillations \cite{tavel1969korotkoff, celler_2021, raftery1968indirect, nitzan2005effects, wilkins1946changes}. This behaviour is shown in (dotted) blue in Fig.\ \ref{fig:pressurevariationBODY}. The pressure on the proximal side of the cuff, the side closer to the heart, is unaffected \cite{raftery1968indirect}, while the pressure on the distal side drops to a constant value, which is $50$ mmHg in our example. This low plateau will be present when the SBP is recorded because the cuff is inflated at the start of the auscultatory measurement, so the artery is closed. Regression analysis of in vivo measurements by Celler et al. \cite{celler2024accurate} showed that the difference between the pressure downstream of the cuff and the systolic pressure accounted for almost 43\% of the variance in systolic measurement error in a cohort of 40 subjects. However, the physics behind this correlation has not been investigated. In parallel with these in vivo measurements, initial tests in our laboratory rig \cite{bassil2024explanation} suggested a relationship between downstream pressure and underestimation. Together, these findings motivate the study presented in this paper to establish a causal link between the lowered downstream pressure and the underestimation of SBP. We have achieved this through controlled, independent variation of the downstream pressure in our laboratory rig, which cannot be produced in vivo. Previous lab-based investigations \cite{sacks_1979, sacksramanandburnell, anlikerandraman} did not observe the underestimation of SBP, because their setup did not reproduce the total closure of the artery and the resulting drop in pressure downstream of the cuff.

\begin{figure}
\centering
\includegraphics[width=8.6cm]{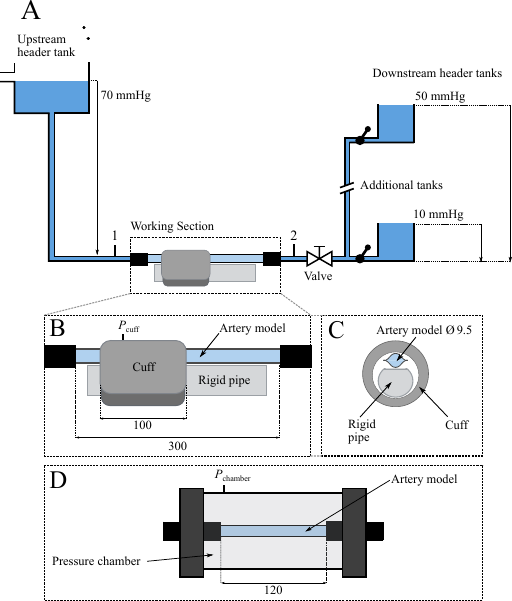}
\caption{A) Schematic of the experimental rig, with upstream and downstream header tanks setting the pressure conditions on either side of the Working Section. The downstream pressure is varied by switching between downstream header tank heights. B) Details of the Working Section of the rig, where the artery model is compressed by a small blood pressure cuff (Omron CD-CS$9$). C) Cross-section of Working Section. D) An alternative Working Section, where the external load is applied to the artery model with a pressure chamber instead of a blood pressure cuff, to provide uniform loading. All lengths in mm.}
\label{fig:experimentalrig}
\end{figure}

\section*{Results}
Reproducing the auscultatory technique in an experimental rig allows us to isolate and measure many variables inaccessible in the body, including upstream and downstream pressures. This controlled environment enables us to systematically and repeatably alter these variables, simplifying the complex in vivo physiology to focus on the physics involved in systolic blood pressure measurement. Although our setup effectively captures the essential physics, it differs from in vivo conditions in aspects such as the material and dimensions of the artery model, and the absolute pressures. Consequently, we avoid direct numerical comparisons with in vivo data, instead emphasising the trends in behaviour observed in our experiments. A schematic of the in vitro rig is shown in Fig.\ \ref{fig:experimentalrig}.

\begin{figure}
\centering
\includegraphics[width=8.7cm]{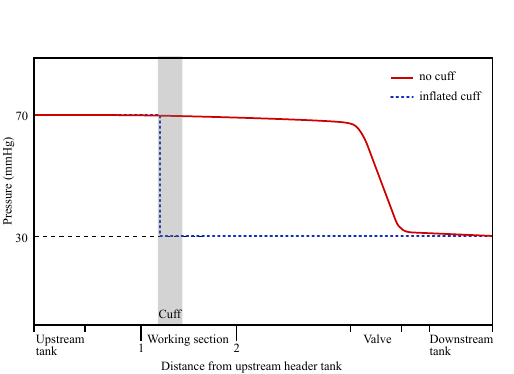}
\caption{Variation in pressure throughout the experimental setup. Labels on the `Distance from upstream header tank' axis correspond to the same labels on the rig schematic in Fig.\ \ref{fig:experimentalrig}.}
\label{fig:pressurevariationRIG}
\end{figure}

Water is used as the working fluid throughout these experiments. Details of this choice are provided in the Supplementary Information: `Using water as the working fluid'. The water flows from the upstream header tank through the artery model in the `Working Section' (labelled in Fig.\ \ref{fig:experimentalrig}) and into one of the downstream tanks, which set the downstream pressure conditions and are interchanged between tests. The pressure variation throughout the rig is shown in Fig.\ \ref{fig:pressurevariationRIG}. As in Fig.\ \ref{fig:pressurevariationBODY}, the red trace shows the pressure variation with no cuff, while the (dotted) blue shows the pressure with the cuff inflated.

Using a constant pressure header tank means that the ‘blood pressure’ measured in the rig is not oscillatory. This results in only a single opening of the artery as the cuff is deflated, rather than the repeated opening and closing seen due to oscillatory pressure in the body. This single opening is equivalent to the first opening of the artery in the body and so represents the systolic measurement. As the systolic reading is the value of interest in this study, with this setup, we maintain the physics that is essential to our investigation while avoiding the complication posed by introducing oscillatory flows. The upstream header tank maintains a fixed pressure of $70$ mmHg at location $1$ ($P_{1}$), upstream of the cuff, irrespective of the flow resistance. Therefore, the pressure does not rise when the cuff is inflated and the artery model closes. This condition is intended to replicate the response in the body, where arterial line measurements show no significant change in upstream blood pressure in response to the closure of the downstream arteries with a cuff \cite{raftery1968indirect}.

An alternative approach, taken in previous experiments \cite{sacks_1979, sacksramanandburnell, anlikerandraman} uses a pump with a parallel route to bypass the closed artery. However, with a parallel route, the upstream pressure will still increase by some fraction when the brachial artery model is closed, due to the change in the total resistance of the network (with resistances acting in parallel, closing one path will lead to a rise in the total resistance). A low resistance bypass, and therefore a high flow rate, would be required to ensure that the change in total resistance when the artery route opened and closed was not significant. The water levels in the upstream and downstream header tanks are more difficult to control with a high flow rate. Critically, using parallel routes makes creating a low downstream plateau challenging because the section distal to the cuff would remain connected to the upstream section via the bypass route, rather than being isolated.

The low pressure distal to the cuff when the artery model is closed (in blue in Fig.\ \ref{fig:pressurevariationRIG}) is possible because, unlike previous setups \cite{sacks_1979, sacksramanandburnell, anlikerandraman} the flow network is not a closed loop, so the inlet and outlet conditions can be set independently. The downstream pressure, at location $2$ ($P_{2}$), is controlled by varying the height of the tanks at the outlet to give pressures between $10$--$50$ mmHg. When the artery model in the Working Section is open (in red in Fig.\ \ref{fig:pressurevariationRIG}), $P_{2} \approx P_{1}$, with only a small drop in pressure along the artery due to the frictional resistance to the flow. This is a good match to the in vivo behaviour, where the average arterial pressure is close to constant along the upper arm (See Fig.\ \ref{fig:pressurevariationBODY}). There is then a large pressure drop between location $2$ and the downstream tank, which is produced with an adjustable valve. In vivo, this pressure drop would happen gradually along the remaining downstream vessels, as shown in Fig.\ \ref{fig:pressurevariationBODY}. The primary role of this valve in the rig is to ensure that, when the artery is open, the majority of the pressure drop occurs across the valve, rather than across the artery model. The valve also reduces the flow rate, which makes it easier to keep the header tank at a constant height. When the artery model is closed, there is no flow and the downstream section is isolated from the upstream header tank. The pressure in the whole section distal to the cuff (measured at location $2$) drops to that of the downstream tank, as shown in blue in Fig.\ \ref{fig:pressurevariationRIG}. The open and closed conditions in the rig replicate those in the body, but with an absence of oscillations in the rig. Although the pressure differences in our rig are smaller than that observed in vivo, of around $50$--$90$ mmHg for a healthy person \cite{tavel1969korotkoff}, they are large enough to reproduce the physical effect of the pressure drop in the body.

In the primary Working Section (Fig.\ \ref{fig:experimentalrig}B and C), the tube representing the artery is compressed against a rigid pipe under an inflatable blood pressure cuff. Previous studies have used rubber models for the artery \cite{sacks_1979, sacksramanandburnell, anlikerandraman}. Rubber tubes are commonly used to investigate the effect of buckling stiffness on overestimation. However, these cylindrical tubes are a poor replica of arteries under larger external pressures. They rapidly increase in stiffness as their cross-sectional area approaches zero and remain partially open at transmural pressures of up to $40$ mmHg \cite{foran2004compression}. In contrast, complete closure of the artery is observed in the body with only a small transmural pressure, close to zero \cite{foran2004compression}. With the incomplete closure of rubber tubes, the absence of flow and the resulting low downstream pressure observed in vivo cannot be recreated.

To overcome these issues, we have used lay-flat tubes to model the artery. Lay-flat tubes have a low resistance to collapse, down to a zero cross-sectional area \cite{geddes2013handbook}, and so exhibit the all-important total closure required to produce the low downstream plateau. Lay-flat tubes also have very low stiffness at the point of buckling \cite{geddes2013handbook}. In this respect, they are dissimilar to brachial arteries. However, buckling stiffness is a common cause of overestimation, and the absence of this stiffness, combined with the lack of surrounding tissue, removes the standard causes of overestimation, making it easier to focus on underestimation. Our tubes have a width of $15$ mm when closed and form an approximate circle of diameter $9.5$ mm when open, compared with a diameter of $4$ mm in vivo \cite{babbs_2015}. Unlike previous studies \cite{sacks_1979,sacksramanandburnell,anlikerandraman}, we are not interested in reproducing the mechanics of buckling of the artery, so we are not concerned about the exact geometry of the tube beyond ensuring that the whole width of the tube can be compressed under the cuff. One disadvantage of lay-flat tubes is their tendency to wrinkle as they close. This wrinkling is hard to predict and affects the closure of the tube. Compressing the tube against a rigid pipe reduces wrinkling and produces a repeatable closure.

Inflatable cuffs apply uneven pressure loading to the artery. To study if this impacts blood pressure measurement, we conducted a series of tests using an air pressure chamber to apply the external pressure instead of a blood pressure cuff. Pressure applied in this way is entirely uniform. This alternative Working Section is shown in Fig.\ \ref{fig:experimentalrig}D. Both the cuff and the pressure chamber are inflated and deflated manually.

\begin{figure}
\centering
\includegraphics[width=8.05cm]{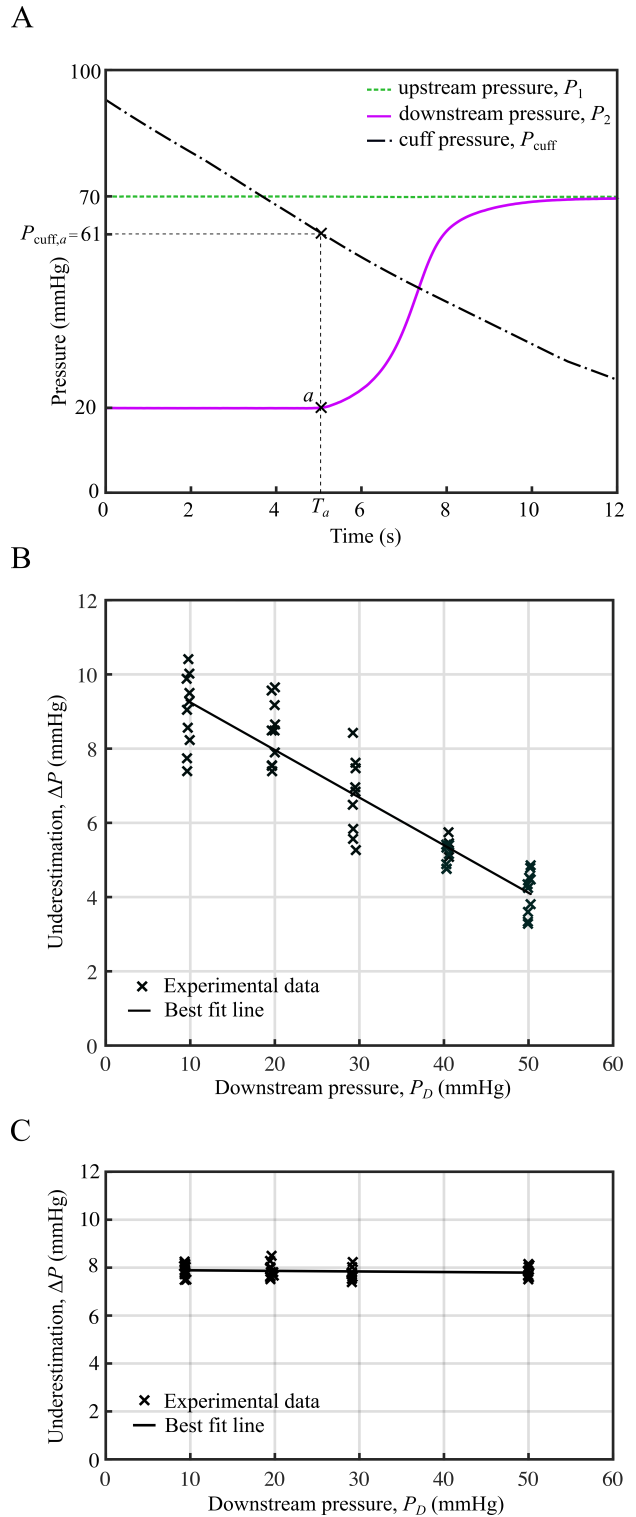}
\caption{A) Pressure traces from a single test run, with a downstream pressure of $20$ mmHg. Measuring the cuff pressure at the time of reopening, $T_{a}$, gives the reading $P_{\mathrm{cuff},a}$ = $61$ mmHg, a $9$ mmHg underestimation of the upstream pressure. B) Underestimation of the upstream pressure, $\Delta P$, measured using a manual blood pressure cuff, for downstream pressures $10$--$50$ mmHg. The equation of the best fit line is $\Delta P = -0.13P_D + 10.53$ C) Underestimation of the upstream pressure measured using a pressure chamber. The equation of the best fit line is $\Delta P = -0.0024P_D + 7.91$. The underestimation is constant across the different downstream pressures tested.}
\label{fig:results}
\end{figure}

\subsection*{Measured cuff pressure}
Fig.\ \ref{fig:results}A depicts an example test run. The cuff is initially inflated to a pressure greater than the upstream pressure and, because the flow through the artery is cut off, the downstream pressure, $P_{2}$, drops to that at the downstream tank ($20$ mmHg in this case). The upstream pressure, $P_{1}$, remains constant at $70$ mmHg, set by the upstream header tank. The cuff around the model artery is then gradually deflated. The rise in the downstream pressure at time $T_{a}$ indicates the opening of the model artery. (Details of selecting the threshold pressure for opening are provided in the Materials and Methods section.) The cuff pressure at this time is recorded ($P_{\mathrm{cuff},a}$). In vivo, this first opening would produce a Korotkoff sound \cite{celler2024accurate}, and the cuff pressure would be recorded as the ‘measured SBP’. In the rig, the cuff pressure is recorded as the ‘measured upstream pressure’, where the upstream pressure represents systolic pressure. The underestimation of the upstream pressure is the difference between the true upstream pressure, $P_{1}$ $=$ $70$ mmHg and the measured upstream pressure, $P_{\mathrm{cuff},a}$ $=$ $61$ mmHg. So, for this test, the underestimation of the upstream pressure is $\Delta P = 9$ mmHg. The test is repeated with different downstream pressures, set by varying the height of the outlet tanks. The underestimation of the upstream pressure is plotted against these downstream pressures in Fig.\ \ref{fig:results}B. There is a clear negative correlation between the downstream pressure and the underestimation of the upstream pressure, with $R = -0.92$ ($p = 1.7 \times 10^{-20}$). Lowering the pressure downstream of the cuff increases the underestimation of the upstream pressure.

\subsection*{Loading with pressure chamber}
Fig.\ \ref{fig:results}C shows the results obtained using the pressure chamber to apply the external loading to the artery model instead of the cuff. Unlike in the cuff case, the downstream pressure does not affect the underestimation of the upstream pressure.

\section*{Discussion}
We have established that low downstream pressure results in the artery model reopening at a lower cuff pressure. We propose the following explanation for this relationship.

\begin{figure}[t]
\centering
\includegraphics[width=8.6cm]{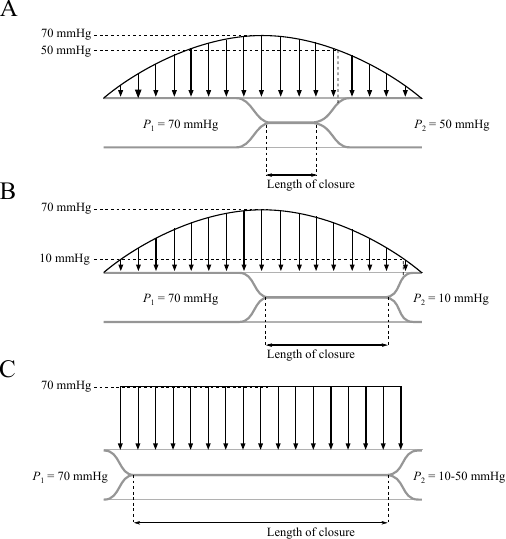}
\caption{A) Closure of tube, shown in grey, under the variable pressure of a cuff reading $70$ mmHg, shown in black, illustrating the length of closure with downstream pressure of $50$ mmHg. B) Closure of the tube under the variable pressure of a cuff reading 70 mmHg, illustrating the length of closure with downstream pressure of $10$ mmHg. C) Closure of the tube under the constant load of an air pressure chamber reading $70$ mmHg. The length of closure remains the same, irrespective of the downstream pressure.}
\label{fig:closurelength}
\end{figure}

Fig.\ \ref{fig:closurelength}A and \ref{fig:closurelength}B show the artery model, hereafter referred to as the tube, compressed by a cuff. The applied cuff pressure of $70$ mmHg is the same in each case, while the downstream pressures are $50$ mmHg and $10$ mmHg, respectively. The pressure applied by the cuff is illustrated in black. Although the cuff pressure reading is $70$ mmHg, the applied pressure will not be constant along the length of the cuff \cite{Deng_Liang, lan_2011, hargens_1987}. The pressure is highest at the centre, with lower pressure towards the edges of the cuff. This means the external pressure applied to the tube exceeds the internal pressure only over a fraction of the cuff length. Therefore, when the tube section central to the cuff is closed, the tube is still open on either side of the centre. The length of the closed section depends on the internal pressure on either side of the point of closure. In Fig.\ \ref{fig:closurelength}A, with an upstream pressure of $70$ mmHg and a downstream pressure of $50$ mmHg, the tube is open on the proximal (upstream) side and closes towards the centre, where the external pressure reaches $70$ mmHg. The tube is then closed until the external pressure is below $50$ mmHg, further down the tube. The internal pressure exceeds the external pressure for the remaining length, so the tube is open. In Fig.\ \ref{fig:closurelength}B, the downstream pressure is decreased to $10$ mmHg. The proximal side of the tube remains open as before, but a longer section is closed, as the tube only reopens where the external pressure drops below $10$ mmHg.

When the cuff is deflated, the tube reopens from the upstream side as the pressure wave travels along the tube. With a greater length of the tube closed, the tube takes longer to fully reopen, so the opening time, $T_{a}$, is later. Therefore, $P_{\mathrm{cuff},a}$ will reach a lower value by this time, resulting in a greater underestimation of the upstream pressure, $\Delta P$. The lower the downstream pressure, the greater the closure length and, therefore, the larger the underestimation.

The results from the pressure chamber tests reinforce this theory. The applied external pressure is uniform across the tube length in the pressure chamber case, as shown in Fig.\ \ref{fig:closurelength}C. There is no decrease in external pressure away from the centre. When the measured external pressure exceeds the upstream pressure, the entire length of the loaded tube will close.  The upstream pressure will still be underestimated because of the time required for the pressure wave to travel along the tube. However, the downstream pressure will not affect this underestimation, since the length of the tube closure is independent of the downstream pressure.

In the rig, the deflation rate of the cuff (or pressure chamber) also influences the underestimation. A more rapid deflation causes the cuff pressure to decrease further in the time required for the tube to fully reopen. (See Fig.\ S$4$ in the Supplementary Material.) To reduce the effect of the deflation rate on our results, for the data shown in Fig.\ \ref{fig:results} we carried out the tests over a narrow range of deflation rates. The influence of the deflation rate on the underestimation data in Fig.\ \ref{fig:results}b is shown in Fig.\ S$5$. The small variation in deflation rate across these tests has no significant impact on the underestimation.

\subsection*{Application to in vivo measurement}

In vivo, when the cuff is inflated above the SBP, the downstream pressure falls to a plateau of $30$--$70$ mmHg \cite{tavel1969korotkoff} compared with a healthy systolic pressure of $120$ mmHg.  With a standard $12$ cm cuff, the length of artery closure under the distal half of the cuff will vary between $0$--$6$ cm, depending on the downstream pressure. The presence of pressure oscillations from the cardiac cycle changes the details of how the length of artery closure affects the level of underestimation, but the principle remains similar. The pressure pulses produced by the heart propagate as waves along the arteries \cite{PNAS_PWV} at the `pulse wave velocity'. With the cuff pressure just below the systolic pressure, the artery will be closed for most of each cardiac cycle. When a systolic pressure peak arrives at the proximal side of the cuff, the artery will start to open as the blood pressure pulse travels along the artery under the cuff. But, the time for which the blood pressure exceeds the cuff pressure, and the pulse can therefore propagate under the cuff, is very small. This time may not be long enough for the pulse to reach the distal edge of the cuff. For example, let us consider a closure of $6$ cm. With a pulse wave velocity of $1$ m/s under the cuff, much slower than in the uncompressed artery \cite{baranger2023fundamental}, the rising front of the pulse wave would take $0.06$ seconds to travel from the proximal side of the closed length to the distal side. For the pulse wave to successfully propagate to the distal side of the cuff, the blood pressure must exceed the cuff pressure for at least this duration of $0.06$ seconds. Otherwise, the blood pressure at the proximal side of the cuff will drop back below the cuff pressure, and the artery will close before the wavefront can reach the distal edge of the cuff. The blood pressure pulse will be unable to propagate, so no Korotkoff sound will be heard.

 \begin{figure}[t]
\centering
\includegraphics[width=8.6cm]{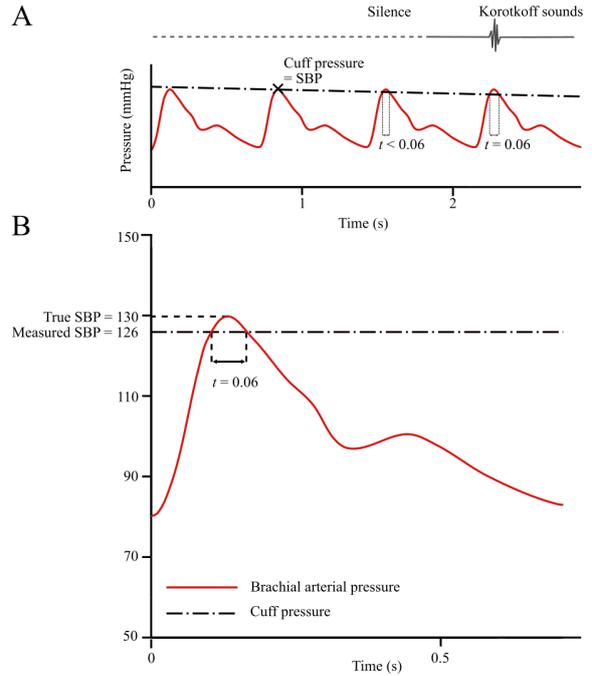}
\caption{An illustration of the underestimation resulting from $6$ cm of closure A) Cuff pressure and arterial blood pressure as the cuff is deflated during an auscultatory measurement. No sounds are heard until the blood pressure exceeds the cuff pressure for $0.06$ seconds. B) For a typical arterial waveform (adapted from \cite{Kroeker_waveform}), the cuff pressure must drop to $126$ mmHg to achieve the $0.06$ seconds required for the artery to open, leading to a $4$ mmHg underestimation.}
\label{fig:arterialwaveform}
\end{figure}

Fig.\ \ref{fig:arterialwaveform} illustrates the level of underestimation that could be caused by $6$ cm of artery closure. Fig.\ \ref{fig:arterialwaveform}A shows representative cuff pressure and arterial blood pressure traces as the cuff is deflated during an auscultatory measurement. No Korotkoff sounds are heard until the blood pressure exceeds the cuff pressure for $0.06$ seconds. This does not occur until the fourth blood pressure pulse, despite the cuff pressure equalling the SBP by the second pulse. In Fig.\ \ref{fig:arterialwaveform}B, we evaluate the underestimation that results from this required $0.06$ seconds for opening. For this standard waveform, reproduced from brachial artery measurements by \cite{Kroeker_waveform}, the SBP will be measured as $126$ mmHg, while the true SBP is $130$ mmHg; a $4$ mmHg underestimation. The level of underestimation depends on the shape of the blood pressure pulse, the pulse wave velocity and the length of closure of the artery, which is governed by the downstream pressure.

The influence of the deflation rate is different in the oscillatory case to the constant upstream pressure case from the rig. The time for the artery to open in vivo is of the order of $0.1$ seconds \cite{baranger2023fundamental}, compared with the order of seconds in the rig. Therefore, even for high deflation rates, there will be very little drop in cuff pressure in the time taken for the artery to open, so the deflation rate will not significantly affect the underestimation caused by the low downstream pressure. However, the deflation rate is still important as it affects the measurement accuracy more generally \cite{Hatsell_phasecycle, zheng_deflrate}. This is why a low deflation rate of $2$--$3$ mmHg/s is recommended for in vivo tests \cite{ogedegbe_2010}. Details of the effect of deflation rate in vivo are provided in the Supplementary Information: `A note on deflation rate'.

Note that the downstream pressure is low only at the start of the auscultatory measurement when the cuff has been inflated to above the SBP, completely occluding the brachial artery. When the DBP is recorded based on the cessation of the Korotkoff sounds, the artery is opening and closing with each cardiac cycle. The distal vessels are, therefore, not isolated from the upstream pressure pulses, and so the downstream pressure is not low \cite{celler_2021}. This cause of underestimation would, therefore, only affect the systolic measurement, in line with the meta-analyses \cite{dankel,picone}.

\subsection*{Conclusions}

We show that, to replicate the underestimation of the systolic blood pressure in a rig, it is essential for the model artery to close completely when the cuff pressure exceeds the systolic pressure, as observed in the body. The closure shuts off the blood supply to the vessels distal to the cuff, causing the pressure in the downstream section of the brachial artery to drop significantly. In our rig, we demonstrate that this pressure drop results in an underestimation of the systolic, but not the diastolic, blood pressure. The lower the downstream pressure, the greater the underestimation of the systolic blood pressure.

Because of the variable pressure applied by the cuff along its length, lowering the downstream pressure results in a greater length of artery closure. A longer closure length, in turn, means more time is required for the artery to open, both in the rig and in vivo, producing greater underestimation. This could be further validated with in vivo ultrasound measurements using a methodology similar to that presented in \cite{baranger2023fundamental}. Understanding the physics underlying the systematic underestimation in cuff-based measurements of systolic blood pressure provides potential avenues for mitigating this error. One possible approach is to apply a calibration factor to these measurements, tailored to individual physiological parameters such as age, height, and arm size. Alternatively, refining the measurement protocol itself may offer a method to enhance accuracy.

\section*{Materials and Methods}
\subsection*{Fabrication of lay-flat tubes}
Lay-flat tubes of width $15$ mm and length $300$ mm were produced by sealing the edges of $440$ gauge polythene film (Transpack) on a strip heater.

\subsection*{Data acquisition and processing}
Pressure sensors ($24$PCCFA$2$G, Honeywell) were used to measure the water pressure at locations $1$ and $2$ in Fig.\ \ref{fig:experimentalrig}. The cuff (and chamber) pressure was measured with a Kulite pressure transducer (XCS--$093$--$5$PSID). The pressure measurements were taken simultaneously, at a frequency of $2$ kHz, using a NI PXIe--$1082$ chassis, with a PXIe--$4499$ analogue input module. The data was post-processed in MATLAB, where calibration factors were applied. The data was digitally filtered using a $4$\textsuperscript{th} order low-pass Butterworth filter with a cut-off frequency of $10$ Hz. This cut-off frequency was chosen to remove noise which could trigger the opening threshold early.

\subsection*{Calibration}
The pressure sensors were calibrated using a mercury manometer. The calibration curves and equations are given in Fig.\ S$1$ and S$2$ in the Supplementary Information. 

\subsection*{Defining the threshold for opening}
In the rig, the opening of the tube is defined by the rising of the downstream pressure from a low plateau, rather than from the presence of Korotkoff sounds. Therefore, there is no discrete value to identify it. Instead, the downstream pressure rises continuously and a threshold value for opening is required. Threshold pressures of $0.03$, $0.2$, $0.5$ and $1.0$ mmHg were tested. The observed relationship between downstream pressure and underestimation of upstream pressure is present in all four cases, with little change in the gradient of the graph of underestimation against downstream pressure (See Fig.\ S$3$). A threshold of $0.2$ mmHg was used throughout.

\section*{Acknowledgements}
We thank Francesca De Domenico, Shiv Kapila and Megan Davies Wykes for their work in the early development of the experimental rig. We are also grateful to Branko Celler for discussing his in vivo underestimation results with us. This work was supported by the Engineering and Physical Sciences Research Council.

\printbibliography

\newpage
\setcounter{figure}{0}
\makeatletter 
\renewcommand{\thefigure}{S\arabic{figure}}
\twocolumn[
  \begin{@twocolumnfalse}
{\Huge Supplementary material}

\subsection*{Using water as the working fluid}
Recreating the non-Newtonian properties of blood is not essential to the investigation of auscultatory measurement, with results obtained using water and glycerol, and steer blood showing good agreement \cite{sacksramanandburnell}. Glycerol is sometimes used to increase the viscosity of water to provide a better match to blood \cite{anlikerandraman}. However, the mechanics of auscultatory measurement have been shown to work over a large range of viscosities (varying by a factor of $30$), covering that of water and blood \cite{anlikerandraman}. While the exact measurements would be influenced by higher viscosity due to the increased resistance to flow, the general trend, lower downstream pressure resulting in greater underestimation, would be unaffected. We have therefore used water as the working fluid throughout our experiments.

\subsection*{A note on deflation rate}
In our experimental rig, the deflation rate is influential because a high deflation rate results in a much larger drop in cuff pressure in the time taken for the artery to open. We demonstrate this by measuring the underestimation over a range of deflation rates, with a fixed downstream pressure of $20$ mmHg. The results from this test are given in Fig.\ S$4$. There is a strong positive correlation between deflation rate and underestimation. In vivo, the time taken for the artery to open is short, of the order of $0.1$ seconds \cite{baranger2023fundamental}, compared with the order of seconds in the rig. Therefore, with a recommended deflation rate of $2$--$3$ mmHg, the cuff pressure would decrease by only $0.2$--$0.3$ mmHg in the time taken for the artery to open. The error caused by the deflation  that occurs while the artery is opening is, therefore, much smaller than the underestimation due to the low downstream pressure. Hence, in vivo, the deflation rate will not significantly affect the underestimation resulting from the low downstream pressure.

However, the deflation rate is still important in auscultatory measurement, as it determines how much the cuff pressure will have decreased from one systolic peak to the next. The resulting source of error is called cardiac cycle phase uncertainty and is explained in detail in \cite{Hatsell_phasecycle}. The key result is that cardiac cycle phase uncertainty will produce an average underestimation of systolic pressure of ``one-half of the per-cardiac-cycle cuff deflation decrement'', and an equal overestimation of the diastolic pressure. With the recommended deflation rate of $2$-$3$ mmHg, cardiac cycle phase uncertainty produces less than $2$ mmHg underestimation of SBP \cite{zheng_deflrate}, compared with an average underestimation of systolic pressure of $5.7$ mmHg found in meta-analysis \cite{picone}. Cardiac phase uncertainty is not significant enough to explain the observed level of systolic pressure underestimation, particularly when considering the known causes of overestimation counteracting the underestimation.

Cardiac cycle phase uncertainty is not relevant to our experimental results, where the upstream pressure is not oscillatory, and, as explained above, does not affect the physics described in the `Application to in vivo measurement' section in the main text. It is included here for completeness and to avoid any uncertainty about the influence of deflation rate.

  \end{@twocolumnfalse}
]

\begin{figure*}[t]
\centering
\includegraphics{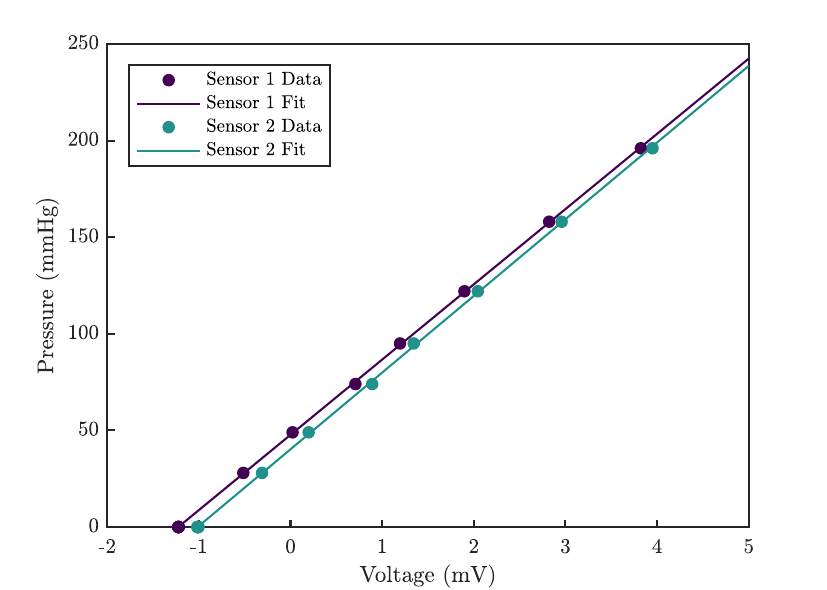}
\caption{Calibration plot for Honeywell sensors. The best fit calibration relationships are  $P_1= 38938.3\times V_1 + 47.7$ and  $P_{2} = 39681.6\times V_{2} + 40.2$, where $V$\textsubscript{1} and $V$\textsubscript{2} are the voltages recorded from the sensors at positions 1 and 2 respectively in Fig.\ 3 in the main text, in V, and $P_1$ and $P_2$ are the pressures recorded on the mercury manometer, in mmHg. Both fits have an $R$\textsuperscript{2} $>$ 0.999.}
\end{figure*}
\clearpage

\begin{figure*}[t]
\centering
\includegraphics{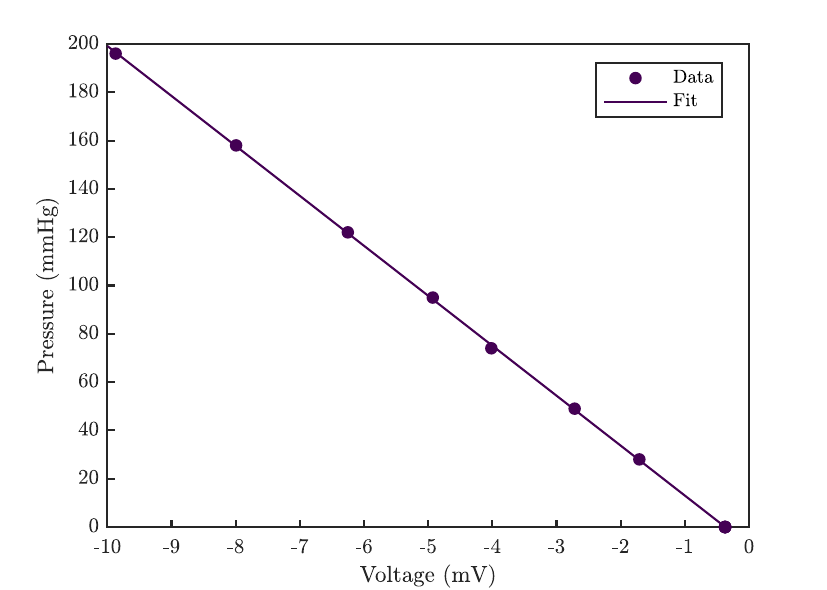}
\caption{Calibration plot for Kulite sensor measuring cuff pressure. The best fit calibration relationship is $P_\mathrm{cuff} = -20682.2 \times V_\mathrm{cuff} + -7.6$ where $V_\mathrm{cuff}$ is the voltage recorded from the Kulite sensor in V, and $P_\mathrm{cuff}$ is the pressure recorded on the mercury manometer, in mmHg. $R$\textsuperscript{2} $>$ $0.999$.}
\end{figure*}
\clearpage

\begin{figure*}[t]
\centering
\includegraphics{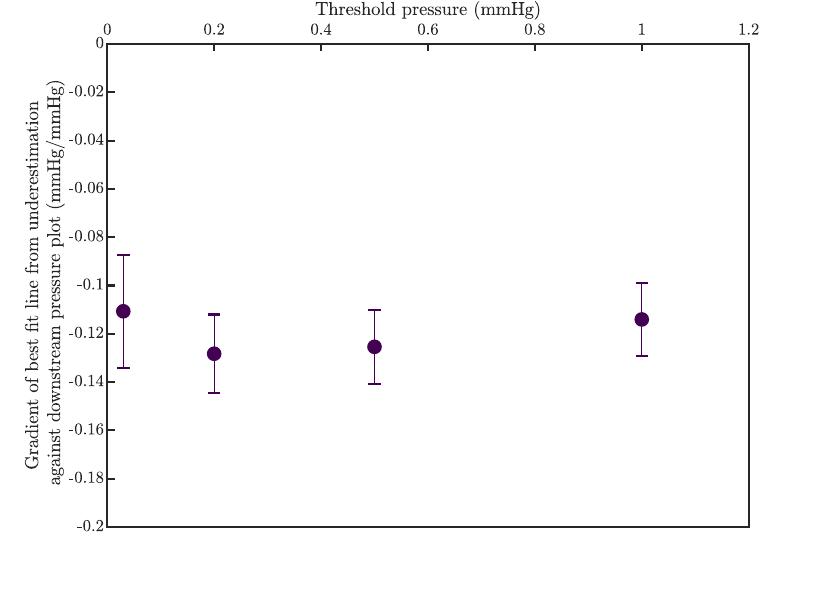}
\caption{The gradient of the underestimation-downstream pressure plots, as in the best fit line in Fig.\ $5$B in the main text, for different threshold pressures. The threshold pressure is the rise in downstream pressure chosen to indicate the opening of the artery. Error bars indicate the $95\%$ confidence interval for each gradient.}
\end{figure*}
\clearpage

\begin{figure*}[t]
\centering
\includegraphics{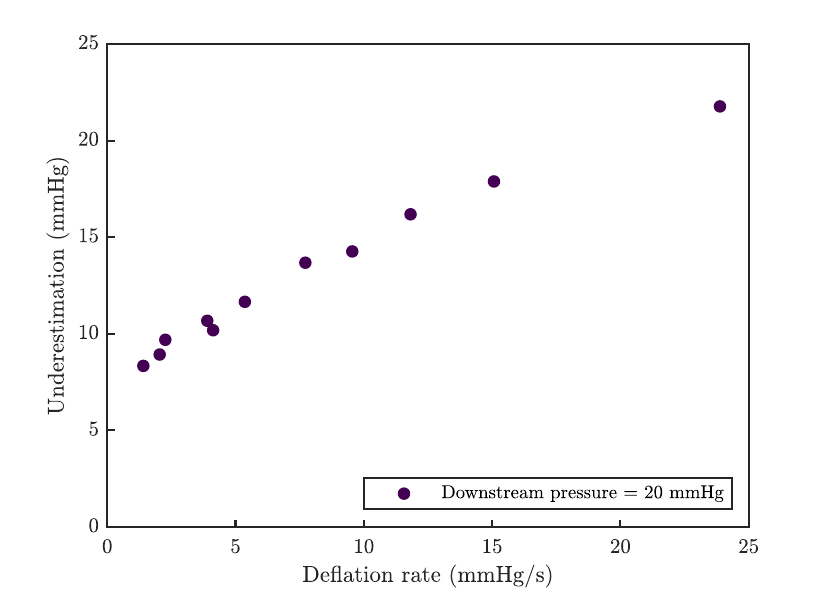}
\caption{Underestimation of upstream pressure against deflation rate. Tests were carried out using the pressure chamber to apply the external load, with a downstream pressure of $20$ mmHg}
\end{figure*}
\clearpage

\begin{figure*}[t]
\centering
\includegraphics{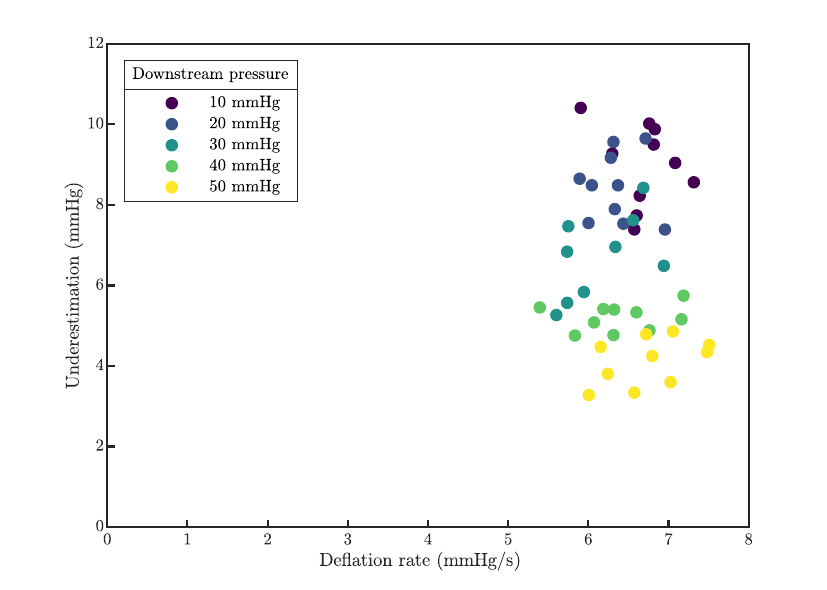}
\caption{Underestimation against deflation rate for the data plotted in Fig.\ $5$B, over five downstream pressures, showing the narrow range of deflation rates over which these tests were carried out.}
\end{figure*}
\clearpage

\end{document}